\documentclass[aps,prl,english,twocolumn,showpacs,preprintnumbers,amsmath,amssymb,floatfix]{revtex4}
\usepackage[T1]{fontenc}
\usepackage[latin9]{inputenc}
\usepackage{graphicx}
\usepackage{amssymb}
\usepackage{babel}


\usepackage{babel}
\makeatother

\begin{document}

\title{Saturation effects in experiments on the thermal Casimir effect}

\author{Bo E. Sernelius}

\affiliation{Division of Theory and Modeling, Department of Physics, Chemistry
and Biology, Link\"{o}ping University, SE-581 83 Link\"{o}ping, Sweden}

\email{bos@ifm.liu.se}

\begin{abstract}
We address three different problematic Casimir experiments in this work. 
The first is the classical Casimir force measured between two metal half
spaces; here in the form of the Casimir pressure measurement between a gold
sphere and a gold plate as performed by Decca et al.  [{\it Phys.  Rev.} D
{\bf 75}, 077101 (2007)]; theory predicts a large negative thermal correction, absent in the
high precision experiment.  The second experiment is the measurement of the
Casimir force between a metal plate and a laser irradiated semiconductor membrane
as performed by Chen et al.  [{\it Phys.  Rev.} B {\bf 76}, 035338 (2007)];
the change in force with laser intensity is larger than predicted by
theory.  The third experiment is the measurement of the Casimir force
between an atom and a wall in the form of the measurement by Obrecht et al. 
[{\it Phys.  Rev.  Lett.} {\bf98}, 063201 (2007)] of the change in
oscillation frequency of a ${}^{87}Rb$ Bose-Einstein condensate trapped to
a fused silica wall; the change is smaller than predicted by theory.  We
show that saturation effects can explain the discrepancies between theory
and experiment observed in all these cases.
\end{abstract}


\maketitle
The Casimir force is very fascinating scientifically and has inspired many
scientists ever since Casimir published his classical paper \cite{Casi} in 1948.  It is
caused by fluctuations in the electromagnetic fields.  What is most
intriguing is the result in the most pure geometry, the one treated by
Casimir himself -- two perfectly reflecting metal plates in vacuum.  Here,
the force is due to true vacuum fluctuations, fluctuations of the
electromagnetic fields in the vacuum surrounding the plates.

The interest in Casimir interactions grew very strong during the last
decade.  This increase in interest was spurred by the torsion pendulum
experiment by Lamoreaux \cite{Lamo}, which produced results with good
enough accuracy for the comparison between theory and experiment to be
feasible.  This stimulated both theorists \cite{BosSer, LamRey, Bordag,
BreAar} and experimentalists \cite{Mohi, DecLop} and the field has grown
constantly since then.  Another reason for this development is the huge
shift of general interest in the science community into nano-science and
nano-technology where these forces become very important.  However, the
field has not been a complete success story.  A dark cloud has been
hovering over this field.  Theory and experiment agree quite well for low
temperatures, but at room temperature, where most experiments are performed
there are serious deviations.  Each new type of experiment has lead to new
puzzling discrepancies between theory and experiment.  Theorists have been
forced to resort to phenomenological approaches to the problems, with new
prescriptions for each new experiment.  This has led to an unfortunate
polarization of the community with those that are content with
phenomenological descriptions on one side and those that want a more
stringent theoretical treatment of the physics on the other.  In this work
we put forward what we think is the solution to the problem or at least the
first step towards a solution.

We address three types of experiment or experimental geometry: Two
interacting metal plates (G1); a semiconductor plate interacting with a metal
plate (G2); an atom interacting with a semiconductor plate (G3).

In all three examples the interaction energy per unit area, $V\left( d
\right)$, can at zero temperature be written on the form \cite{Ser}
\begin{equation}
 V\left( d \right) = \frac{\hbar }{\Omega }\sum\limits_{\bf{k}}
{\int\limits_0^\infty {\frac{{d\omega }}{{2\pi }}} } \ln \left[ {f\left(
{k,i\omega } \right)} \right],
\label{equ1}
\end{equation}
where $d$ is the distance between the objects, {\bf k} the two-dimensional
wave vector in the plane of the plate(s), $\Omega $
 the area of a plate, and
$f\left( {k,\omega } \right) = 0$ is the condition for an electromagnetic
normal mode in the particular geometry.  The integration is along the imaginary
frequency axis.  At finite temperature the integration is replaced by a
discrete summation over Matsubara frequencies,
\begin{equation}
V\left( d \right) = \frac{1}{{\beta \Omega }}\sum\limits_{\bf{k}}
{\sum\limits_{\omega _n }  }^{'} \ln \left[ {f\left( {k,i\omega _n } \right)}
\right];\;\omega _n = \frac{{2\pi n}}{{\hbar \beta }}.
\label{equ2}
\end{equation}
Alternatively one may integrate along the real frequency axis,
\begin{equation}
V\left( d \right) = \frac{{2\hbar }}{\Omega }\sum\limits_{\bf{k}}
{{\mathop{\rm Im}\nolimits} \int\limits_0^\infty {\frac{{d\omega }}{{2\pi
}}} } \left[ {n\left( \omega \right) + {1 \mathord{\left/ {\vphantom {1 2}}
\right.  \kern-\nulldelimiterspace} 2}} \right]\ln \left[ {f\left(
{k,\omega } \right)} \right],
\label{equ3}
\end{equation}
where $n\left( \omega \right) = \left[ {\exp \left( {\hbar \beta \omega }
\right) - 1} \right]^{ - 1} $ is the distribution function for massless
bosons.  This form can also be used at zero temperature; then the
distribution function vanishes.

 The force per unit area, or pressure, is obtained as the derivative with
 respect to distance, $F\left( d \right) = - {{dV\left( d \right)}
 \mathord{\left/ {\vphantom {{dV\left( d \right)} {dd}}} \right. 
 \kern-\nulldelimiterspace} {dd}}$.  In all three geometries there are two
 groups of normal mode, transverse magnetic (TM) and transverse electric
 (TE), each with a different mode condition function.  The interaction
 potential is a sum of two terms, $ V\left( d \right) = V^{TM} \left( d
 \right) + V^{TE} \left( d \right) $.  In G1 and G2 the mode
 condition functions are
\begin{equation}
  f^{\scriptstyle TM, \hfill \atop \scriptstyle TE \hfill} \left( {k,\omega
  } \right) = 1 - e^{ - 2\gamma _0 \left( {k,\omega } \right)d}
  r_{01}^{\scriptstyle TM, \hfill \atop \scriptstyle TE \hfill} \left(
  {k,\omega } \right)r_{02}^{\scriptstyle TM, \hfill \atop \scriptstyle TE
  \hfill} \left( {k,\omega } \right),
\end{equation}
where the Fresnel amplitude reflection coefficients at an interface between 
medium {\it i} and {\it j} are
\begin{equation}
r_{ij}^{TM} \left( {k,\omega } \right) = \frac{{\varepsilon _j \left(
\omega \right)\gamma _i \left( {k,\omega } \right) - \varepsilon _i \left(
\omega \right)\gamma _j \left( {k,\omega } \right)}}{{\varepsilon _j \left(
\omega \right)\gamma _i \left( {k,\omega } \right) + \varepsilon _i \left(
\omega \right)\gamma _j \left( {k,\omega } \right)}},
\end{equation}
for TM modes (p-polarized waves) and
\begin{equation}
r_{ij}^{TE} \left( {k,\omega } \right) = \frac{{\gamma _i \left( {k,\omega
} \right) - \gamma _j \left( {k,\omega } \right)}}{{\gamma _i \left(
{k,\omega } \right) + \gamma _j \left( {k,\omega } \right)}},
\end{equation}
for TE modes (s-polarized waves), respectively.  We have let the objects be of medium 1
and 2 and let the surrounding vacuum be denoted by medium 0.  The gamma
functions are
\begin{equation}
\gamma _j \left( {k,\omega } \right) = \sqrt {k^2 - \varepsilon _j \left(
\omega \right)\left( {{\omega \mathord{\left/
 {\vphantom {\omega  c}} \right.
 \kern-\nulldelimiterspace} c}} \right)^2 } ;\;j = 0,1,2.
\end{equation}

\begin{figure}
\includegraphics[width=8cm]{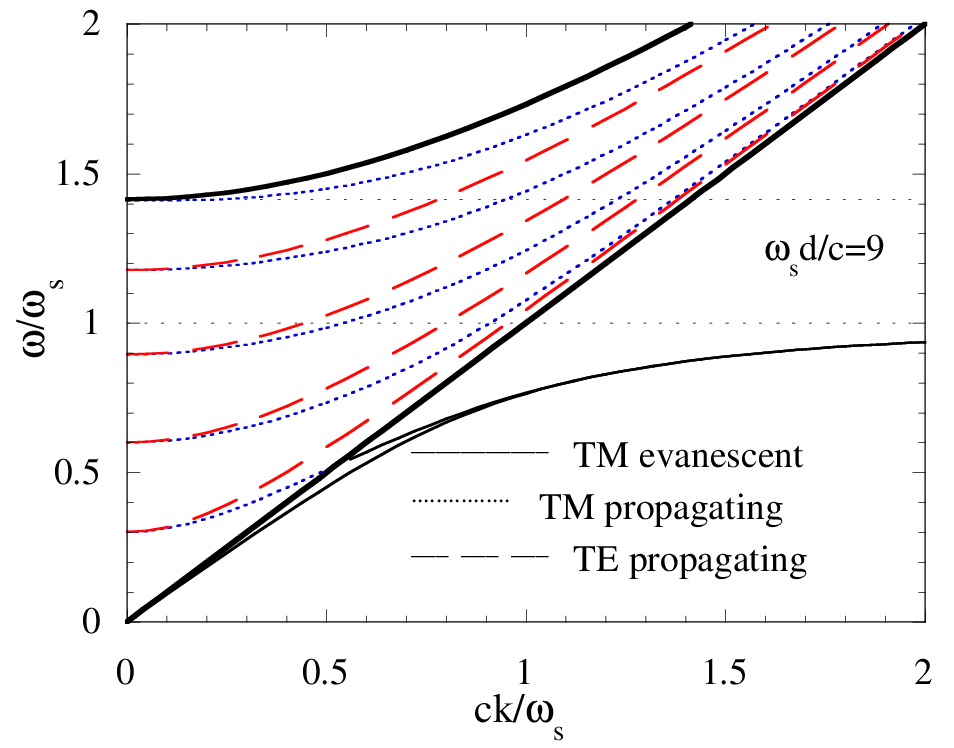}
\caption{Dispersion curves for the modes between two gold plates in 
absence of dissipation.  The frequencies are in units of $\omega _s $, the
surface plasmon frequency. The solid straight line is the light dispersion
curve in vacuum; the dashed (dotted) curves are TE (TM) propagating modes;
the thin solid curves are evanescent TM modes; the thick solid curve is the
lower boundary for transverse bulk modes in the plates.  From Ref. 
\cite{Ser2}}
\label{figu1}
\end{figure}
\begin{figure}
\includegraphics[width=8cm]{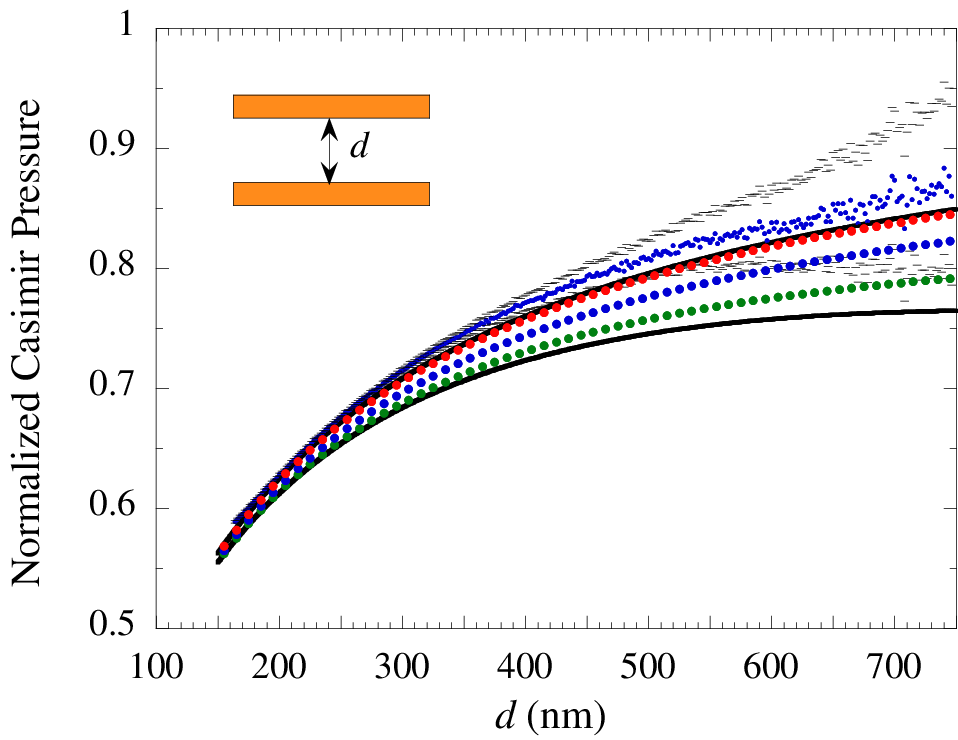}
\caption{Casimir pressure between two gold plates. The experimental result 
from Ref. \cite{Decca} is shown as dots and the endpoints of the error bars 
are indicated by horizontal bars; the upper (lower) solid curve is the 
traditional theoretical zero (Room) temperature result; the circles are 
the present results with damping parameters 0.01, 0.1, and 1.0, 
respectively counting from below.}
\label{figu2}
\end{figure}

Let us now study the dispersion curves for the electromagnetic normal modes
in G1 shown in Fig. \ref{figu1} for two gold plates \cite{Ser2}.  This
figure is valid in neglect of dissipation in the plate materials.  The
modes are propagating (evanescent) above and to the left (below and to the
right) of the light dispersion curve.  Note that there are no TE evanescent
modes.  When the system is allowed to have dissipation there are modes
everywhere.  Each original mode is replaced by a continuum of modes
\cite{Ser3}.  Evanescent TE modes appear and the continuum extends all the
way down to the momentum axis.  These modes are the cause of all the
problems with the thermal Casimir force in this geometry.  The proposed
prescription has been to neglect the dissipation in the intraband part of
the dielectric function but keep it in the interband part \cite{MosGey}. 
The experimental result \cite{Decca} for the normalized Casimir pressure at
295 K is shown as dots in Fig.  \ref{figu2}.  The bars are the endpoints of
the experimental error bars.  The upper (lower) thick solid curve is the
theoretical result for zero temperature (295 K) calculated with Eqs. 
(\ref{equ1}) and (\ref{equ2}), respectively.  We note that the zero
temperature result agrees much better with the experimental result.  The
large negative thermal correction comes entirely from the TE evanescent
modes \cite{TorLam}.  We will demonstrate this in more detail in a
forthcoming publication \cite{Ser4}.  All curves are normalized with the
zero temperature Casimir pressure between two perfect metal plates,
${{\hbar c\pi ^2 } \mathord{\left/ {\vphantom {{\hbar c\pi ^2 } {\left(
{240z^4 } \right)}}} \right.  \kern-\nulldelimiterspace} {\left( {240z^4 }
\right)}}$.  We have neglected surface roughness effects. 

Let us now explain our view of what goes wrong in the theory of the thermal
Casimir effect in presence of dissipation.  The traditional theory relies
fully on the concept of electromagnetic normal modes.  These are assumed to
be independent massless bosons.  The possibility to excite one of these
modes is assumed to be completely independent of how many modes are already
excited.  An excitation of a mode involves excitations of the charged
particles in the system, electrons in the geometries studied here.  These
are the sources of the fields.  Now, the electrons are fermions and there
is at most one electron in each particle state.  An electron that is
excited at one instant of time cannot be excited again -- the state is
empty.  The more modes that are excited the more difficult it is to excite
new modes --- there are saturation effects.  In the theoretical treatment
this is not taken care of.  In most cases this fact will not cause any
problems, but sometimes it could.  We think that the thermal Casimir effect
is one such case.  When dissipation is included each mode is replaced by a
continuum of an infinite number (for an infinite system) of new modes.  The
distribution function diverges towards zero frequency and the saturation
effects should appear here.  This is very difficult to treat in a strict
way.  We use an approximation which is very easy to implement.  We shift
the distribution function in Eq.  (\ref{equ3}) downwards in frequency, so
that it never reaches the point of divergence, by adding a damping
parameter, $D$,
\begin{equation}
 \tilde n\left( \omega \right) = \left[ {\exp \left( {\hbar \beta \omega +
 D} \right) - 1} \right]^{ - 1}.
\end{equation}
 The discrete frequency summation in Eq.  (\ref{equ2}) is the result of the
 poles of the distribution function that all fall on the imaginary axis,
 see Ref.  \cite{Ser}.  Our new distribution function has its poles shifted
 away from the axis the distance ${D \mathord{\left/ {\vphantom {D {\hbar
 \beta }}} \right.  \kern-\nulldelimiterspace} {\hbar \beta }}$ into the
 left half plane. The new form is
\begin{equation}
V\left( d \right) = \frac{1}{{\beta \Omega }}\sum\limits_{\bf{k}}
{\sum\limits_{\omega _n } }^{' } \frac{1}{\pi }\int\limits_{ - \infty }^\infty
{\frac{{\left( {{D \mathord{\left/ {\vphantom {D \beta }} \right. 
\kern-\nulldelimiterspace} \beta }} \right)\ln \left[ {f\left( {k,i\omega
'} \right)} \right]}}{{\left( {\omega ' - \omega _n } \right)^2 + \left(
{{D \mathord{\left/ {\vphantom {D \beta }} \right. 
\kern-\nulldelimiterspace} \beta }} \right)^2 }}}. 
\label{equ9}
\end{equation}

\begin{figure}
\includegraphics[width=8cm]{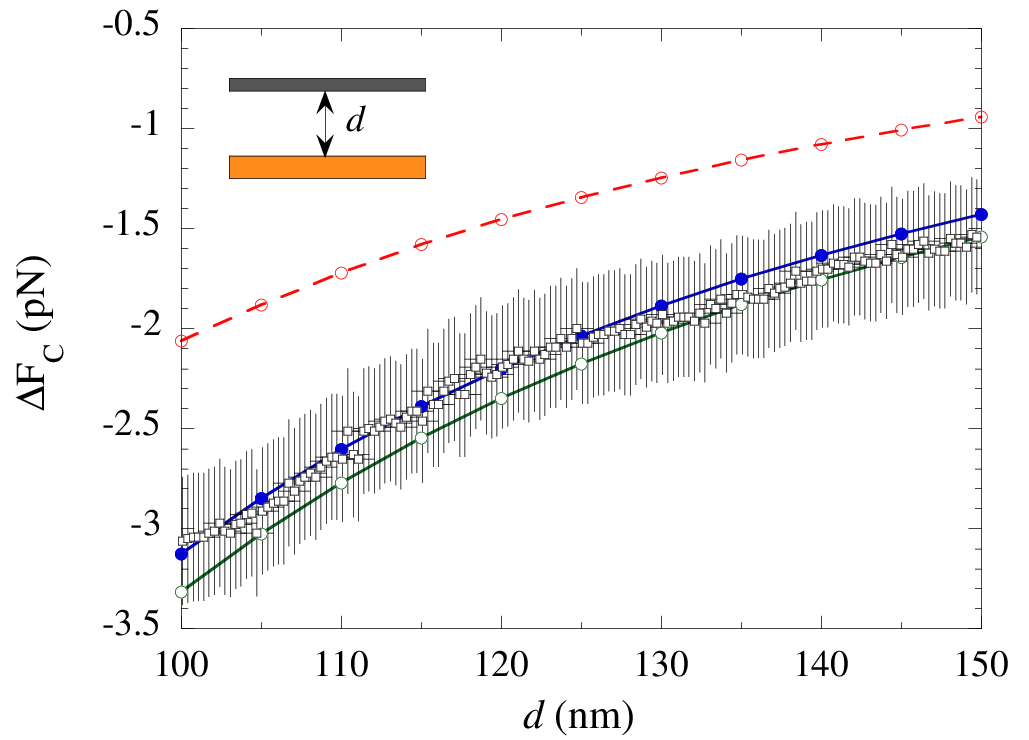}
\caption{The change in Casimir force, at 300 K, between a gold sphere and a silicon 
membrane with and without laser irradiation.  The open squares with error
bars are the experimental \cite{Chen} result.  The dashed curve with
open circles is the theoretical result without saturation effects.  The solid curve
with filled (open) circles is our present result with $D$ equal to 0.01
(0.1).  }
\label{figu3}
\end{figure}
Each term in the summation is replaced by an integral. for small $D$ 
values it is enough to replace only the zero frequency term. We will 
expand on this in Ref. \cite{Ser4}. The circles in Fig. (\ref{figu2}) are 
the results with damping parameters 0.01, 0.1, and 1.0, 
respectively counting from below. We have used Eq. (\ref{equ3}) with the 
modified distribution function to get the thermal correction.

The second geometry, G2, is a gold plate and a laser irradiated
semiconductor membrane as performed by Chen et al.  \cite{Chen}.  They
measured the change in force with the laser irradiation compared to without
any irradiation.  The results are shown in Fig.  \ref{figu3}.  The open
squares with error bars are the experimental result.  The dashed curve with
open circles is the theoretical result for 300 K. The deviations are clear. 
In this geometry it is not enough to neglect dissipation to get agreement
with experiment.  Besides, it is now the TM modes that cause the problems. 
One postulated that for the non-irradiated semiconductor one should
completely omit the contribution to the dielectric function from the
thermally excited carriers.  That brought the theory and experiment into
agreement, see Fig.  10 in Ref.  \cite{Chen}.  The solid curve with filled
(open) circles is our saturation based result with $D$ equal to 0.01 (0.1). 
Here we have used Eq.  (\ref{equ2}) and just modified the zero frequency
contribution according to Eq.  (\ref{equ9}).

\begin{figure}
\includegraphics[width=8cm]{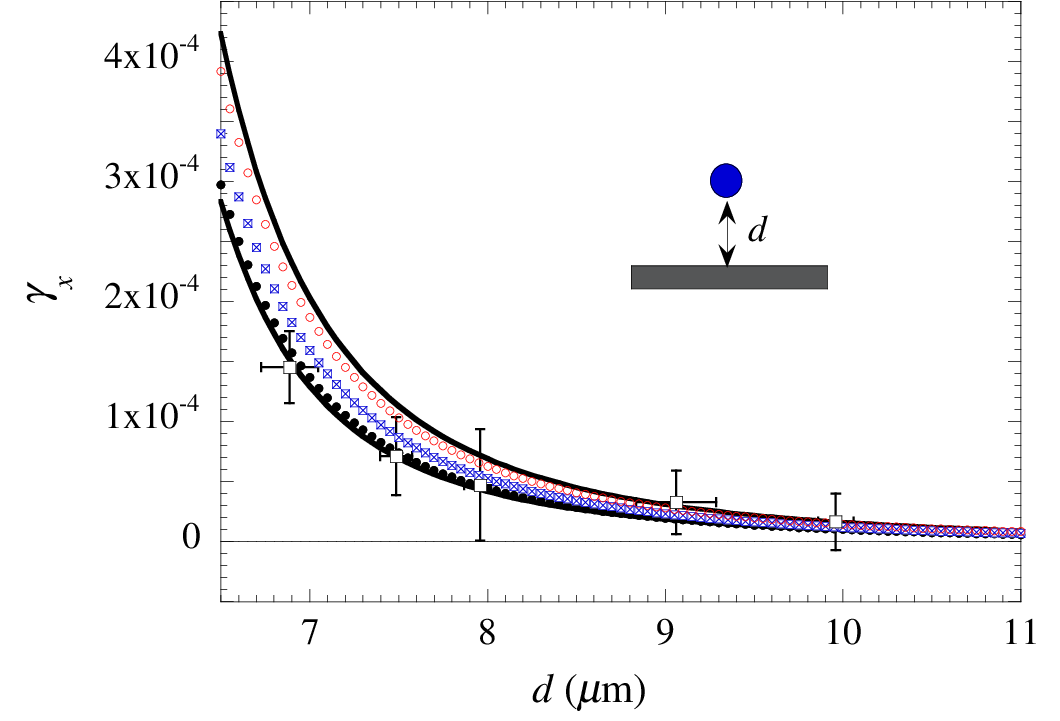}
\caption{Fractional change in trap frequency for a Rb atom near a silica 
wall versus separation in thermal equilibrium.  The open squares are the
experimental result \cite{Obrecht}.  The upper (lower) curve is the theoretical result
including (neglecting) the conductivity from the few thermal carriers in the
silica wall.  The circles are our present results for the $D$ values $10^{
- 10}$, $10^{ - 11}$, and $10^{ - 12}$, respectively, counted from below.}
\label{figu4}
\end{figure}

The third geometry, G3, is a Rb atom near a fused silica wall.  We
study the fractional change in trap frequency versus separation when both
the surroundings and the wall have the same temperature, 310 K.  In Fig. 
\ref{figu4} the experimental result \cite{Obrecht} is shown as open squares with 
error bars.  We use the same formalism as in Ref.  \cite{KliMos} to find
the relation between the change in trap frequency and the Casimir force. 
The upper (lower) curve is the theoretical result, without saturation,
including (neglecting) the conductivity from the few thermal carriers in
the silica wall.  We see that also here the neglect of the contribution, to
the dielectric function of the silica wall, from the very few thermally
excited carriers brings the theoretical result into agreement with
experiment.  This neglect is the postulated remedy in Ref.
\cite{KliMos}.  In this geometry, just as in G2, the TM modes cause the
problems and it is not enough to neglect dissipation to get good agreement
between theory and experiment.  To include saturation effects we have
used Eq.  (\ref{equ2}) and just modified the zero frequency contribution
according to Eq.  (\ref{equ9}).  We note that in this experiment it is
enough to have a damping parameter as small as $10^{ - 10}$ to bring the
theoretical result into agreement with experiment.  We have assumed that
the thermally excited carriers in the wall material have the conductivity
$100\,s^{ - 1}$ ($ \sim 10^{ - 10} {\rm{ohm}}^{ - 1} {\rm{cm}}^{ - 1}$),
which is the upper limit of the range given in Ref.  \cite{KliMos}.  Using
smaller values leads to even weaker demands on the damping parameter.  

In summary we have proposed that saturation effects are responsible for the
discrepancy between theory and experiment in several quite different
Casimir geometries.  We have treated saturation within a very simple
calculation model and demonstrated that the problems may go away in all
cases.  Other very recent theoretical models \cite{Pita, DalLam} have been
proposed for the treatment of dielectric materials with a very small amount
of free carriers, applicable to the G2 and G3 geometries.  However, no
quantitative comparison with the experiments has been presented.

We are grateful to R.S. Decca, G.L. Klimchitskaya, and U. Mohideen for
providing us with experimental data.  The research was sponsored by the
VR-contract No:70529001 and support from the VR Linn\'{e} Centre LiLi-NFM
and from CTS is gratefully acknowledged.


\begin{thebibliography}{10}



\bibitem{Casi} H. B. G. Casimir, {\it Proc. K. Ned. Akad. Wet.} {\bf 51}, 793
(1948).

\bibitem{Lamo} S. K. Lamoreaux, {\it Phys.
Rev. Lett. } {\bf 78}, 5 (1997).

\bibitem{BosSer}M. Bostr{\"om} and Bo E. Sernelius, {\it Phys. Rev. Lett} {\bf 
84}, 4757 (2000).

\bibitem{LamRey} A. Lambrecht, and S. Reynaud, {\it Eur. Phys. J.} D {\bf 
8}, 309 (2000).

\bibitem{Bordag} M. Bordag, B. Geyer, G. L. Klimchitskaya, and V. M.
Mostepanenko {\it Phys.  Rev.  Lett} {\bf 87}, 259102 (2001).

\bibitem{BreAar} I. Brevik, J. B. Aarseth, and J. S. H{\o}ye, {\it Phys. 
Rev.} E {\bf 66}, 026119 (2002).

\bibitem{Mohi} U. Mohideen, and A. Roy, {\it Phys.
Rev. Lett. } {\bf 81}, 4549 (1998).

\bibitem{DecLop} R. S. Decca, D. L{\'o}pez, E. Fischbach, and D. E. 
Krause, {\it Phys. Rev. Lett.} {\bf 91}, 050402 (2003).



\bibitem{Ser} Bo E. Sernelius, {\it Surface Modes in Physics} (Wiley-VCH, 
Berlin, 2001).



\bibitem{Ser2} Bo E. Sernelius, {\it Phys. Rev.} B {\bf 71}, 235114 (2005).

\bibitem{Ser3} Bo E. Sernelius, {\it Phys. Rev.} B {\bf 74}, 233103 (2006).

\bibitem{MosGey}V. M. Mostepanenko and B. Geyer, {\it J. Phys. 
A:Math. Theor.} {\bf 41}, 164014 (2008).

\bibitem{Decca} R. S. Decca, D. Lop\'{e}z, E. Fischbach, G. L. 
Klimchitskaya, D. E. Krause, and V. M. Mostepanenko, {\it Phys. Rev.} D 
{\bf 75}, 077101 (2007).

\bibitem{TorLam} J. R. Torgerson, and S. K. Lamoreaux, {\it Phys. Rev.} E 
{\bf 70}, 047102 (2004).



\bibitem{Ser4}Bo E. Sernelius, to be published

\bibitem{Chen} F. Chen, G. L. Klimchitskaya, V. M. Mostepanenko, and U. 
Mohideen, {\it Phys. Rev.} B {\bf 76}, 035338 (2007).

\bibitem{Obrecht} J. M. Obrecht, R. J. Wild, M. Antezza, L. P. Pitaevskii, 
S. Stringari, and E. A. Cornell, {\it Phys. Rev. Lett.} {\bf98}, 
063201 (2007).

\bibitem{KliMos} G. L. Klimchitskaya and V. M. Mostepanenko, {\it J. Phys. 
A:Math. Theor.} {\bf 41}, 312002 (2008).





\bibitem{Pita}L. P. Pitaevskii, {\it Phys.
Rev.  Lett.  } {\bf 101}, 163202 (2008).


\bibitem{DalLam} D. A. R. Dalvit and S. K. Lamoreaux, {\it Phys.
Rev. Lett. } {\bf 101}, 163203 (2008).

















\end{thebibliography}
\end{document}